# Misalignment between the Directions of Propagation and Decay of Nanoscale-confined Polaritons

*Kirill V. Voronin[1,2], Gonzalo Álvarez-Pérez[3], Aitana Tarazaga Martín-Luengo[4,5], Pablo Alonso-González[4,5,†], Alexey Y. Nikitin[1,6,‡]*

[1]*Donostia International Physics Center (DIPC), Donostia-San Sebastián 20018, Spain*

[2]*Universidad del Pais Vasco/Euskal Herriko Unibertsitatea, 20080 Donostia/San Sebastián, Basque Country, Spain*

[3]*Istituto Italiano di Tecnologia, Center for Biomolecular Nanotechnologies, 73010 Arnesano, Italy*

[4]*Department of Physics, University of Oviedo, Oviedo 33006, Spain*

[5]*Center of Research on Nanomaterials and Nanotechnology, CINN (CSIC-Universidad de Oviedo), El Entrego 33940, Spain*

[6]*IKERBASQUE, Basque Foundation for Science, Bilbao 48013, Spain*

ABSTRACT: Anisotropic van der Waals crystals have gained significant attention in nanooptics and optoelectronics due to their unconventional optical properties, including anomalous reflection, canalization, and nanofocusing. Polaritons—light coupled to matter excitations—govern these effects, with their complex wavevector encoding key parameters such as wavelength, lifetime, field confinement, and propagation direction. However, determining the complex wavevector, particularly the misalignment between its real and imaginary parts, has remained a challenge due to the complexity of the dispersion relation. Here, using near-field nanoimaging, we introduce a self-consistent method to extract the complex wavevector from polaritonic near-field images. We experimentally reveal a strong misalignment between the real and imaginary components of the wavevector, significantly impacting the interpretation of near-field experiments. Our findings establish a new paradigm for optical nanoimaging, providing a robust framework for accurately extracting polariton parameters and advancing the broader field of nanooptics of lossy anisotropic crystals.

When imagining an electromagnetic wave in a medium, we usually assume its amplitude decays along the propagation direction. This assumption indeed holds for *homogeneous waves* in lossy isotropic media. [1] By contrast, evanescent waves decay even without absorption, with attenuation perpendicular to their wavefronts, enabling deep subwavelength transverse confinement—a key driver for late-20th-century nanooptics. A prime example is surface polaritons — hybrid waves from strong coupling of electromagnetic fields with collective dipole excitations. [2-6] Their fields decay perpendicular to an interface, confining them near boundaries between media of opposite permittivities and enabling subwavelength resolution in various near-field microscopy techniques.

Mathematically, wave propagation and spatial decay are described by an exponential function, $\exp(i\mathbf{k}\mathbf{r})$, with a complex wavevector $\mathbf{k} = \mathbf{k}' + i\mathbf{k}''$. Its real part, $\mathbf{k}'$ sets the direction and velocity of wavefronts (phase velocity), while its imaginary part, $\mathbf{k}''$, governs amplitude decay. For homogeneous waves in isotropic media, $\mathbf{k}'$ and $\mathbf{k}''$ are parallel, while for evanescent waves in isotropic and lossless media, they are perpendicular, $\mathbf{k}' \cdot \mathbf{k}'' = 0$, as dictated by the standard dispersion relation $\mathbf{k}^2 = \varepsilon\omega^2/c^2$ (Figure 1a, b). In contrast, in anisotropic or lossy materials, the lack of symmetry can tilt the decay direction relative to phase propagation, making $\mathbf{k}'$ and $\mathbf{k}''$ neither parallel nor strictly orthogonal. This is related to the tilt of the wavefronts relative to the energy propagation direction in near-field nanoimaging of polaritons in strongly anisotropic media. [7, 8] In particular, anisotropic van der Waals (vdW) crystals (such as α-MoO$_3$, [9] α-V$_2$O$_5$, [10,]

[11] WTe$_2$, [12] MoOCl$_2$, [13, 14] β-Ga$_2$O$_3$, [15] CdWO$_4$[16] and calcite[17]) have recently received considerable attention, as they can exhibit highly directional propagation, [9, 12, 18] canalization, [11, 19-25] anomalous reflection, [7] refraction, [8] tunability, [18, 26, 27] ghost and leaky waves, [13, 17, 28, 29] shear-effects, [16, 30] and exceptionally strong confinement. [31]

Typically, these phenomena are probed using techniques like near-field microscopy, which is widely employed to extract both the polaritonic wavelength and propagation length, thereby enabling direct reconstruction of their complex-valued wavevector **k** as a function of frequency. [9, 32-36] Crucially, although polaritons primarily reside within an anisotropic or lossy material, part of their field extends into isotropic media above or below, commonly into air, where most nanoimaging measurements are performed. Thus, in the medium where the measurement is carried out, the amplitude and phase of the evanescent wave evolve orthogonally, still fulfilling **k**′ ⊥ **k**″. At the same time, the projections of the wavevector onto the sample face, i.e., in-plane real and imaginary components of the wavevector, **k**$_∥$′ and **k**$_∥$″, respectively, can form arbitrary angles. Figure 1c depicts the distribution of the free-space electric field ($E_z$ component, without loss of generality) for an electromagnetic wave propagating along the surface of an anisotropic layer. The arrows indicate the total wavevector components in vacuum, **k**′ ⊥ **k**″, as well as their in-plane projections **k**$_∥$′ ∦ **k**$_∥$″. Note, inside a layer of anisotropic material, the condition **k**′ ⊥ **k**″ does not hold, and the relative orientation of **k**′ and **k**″ can be arbitrary. In particular, for surface-confined polaritons, $k_z$ is predominantly imaginary, whereas for volume-confined polaritons, it is real. Owing to its general formulation, the present approach remains applies to both. The misalignment between **k**$_∥$′ and **k**$_∥$″ is not only of particular importance for extracting experimentally key polariton parameters in anisotropic materials but also a very fundamental property in electromagnetics, related to the energy and momentum of a wave. However, it has been largely overlooked up to now.

In this Letter, we experimentally prove that the wavefronts and the decay direction of polaritons in strongly anisotropic media exhibit a tilt. To do so, we introduce a theoretical method for calculating both the real ($\mathbf{k}'$) and imaginary ($\mathbf{k}''$) components of the polariton wavevector in an arbitrary medium, allowing us to disentangle them in near-field measurements. Furthermore, we prove that the group velocity of highly confined in-plane anisotropic polaritons is parallel to the Poynting vector—the time-averaged directional energy flux of the electromagnetic wave— and equal to the energy transfer velocity in the low-loss frequency range. Our findings provide a deeper understanding of the fundamental properties of anisotropic polaritons and enable the correct extraction of their parameters, such as e.g., the direction-dependent lifetime.

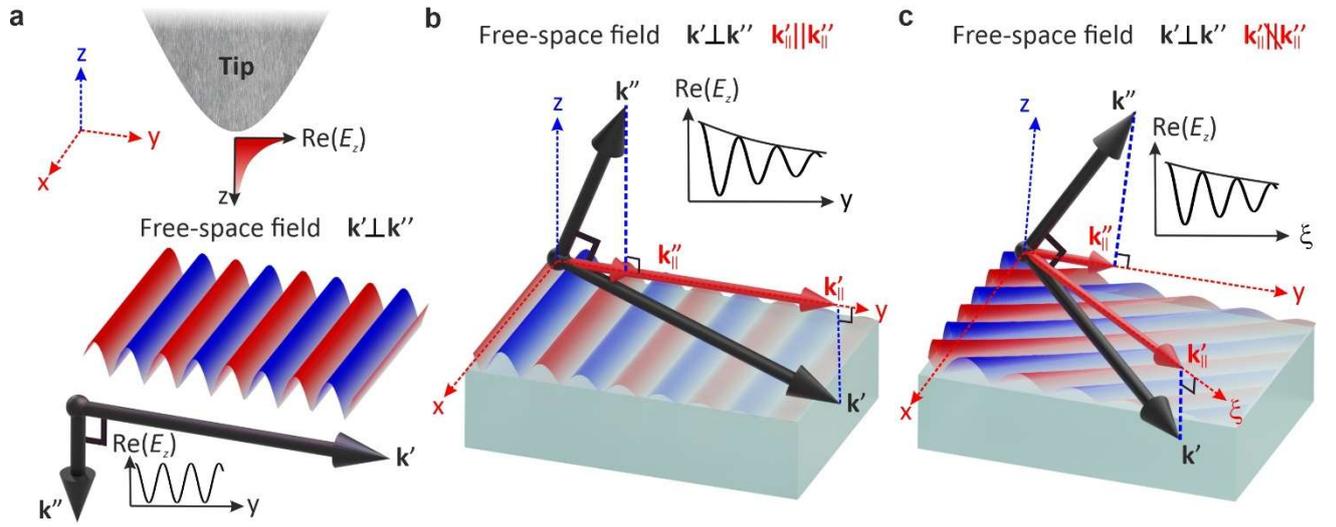

**Figure 1.** Examples of evanescent plane waves with non-collinear $\mathbf{k}'$ and $\mathbf{k}''$. (a) One of the in-plane Fourier harmonics of the near field created by a tip. (b) Surface wave at the interface of a lossy in-plane isotropic material. (c) Surface wave at the interface of a lossy in-plane anisotropic material. In (b, c) $k'_z$ takes negative values; the thick red arrows show the in-plane wavevectors $\mathbf{k}'_{\parallel}$ and $\mathbf{k}''_{\parallel}$.

We consider the vdW polar crystal α-MoO$_3$ as a representative highly anisotropic biaxial material supporting low-loss in-plane anisotropic phonon polaritons (PhPs). [9] This material exhibits several Reststrahlen bands (RBs) in the mid- and far-infrared spectral regions in which at least one diagonal component of the Re($\hat{\varepsilon}$) tensor becomes negative.[37, 38] Within the different RBs, α-MoO$_3$ supports either in-plane elliptic or hyperbolic polaritons, whose isofrequency curves (IFCs) have elliptic-like or hyperbolic-like shapes, respectively, depending on the sign of the in-plane components of the Re($\hat{\varepsilon}$) tensor.[9, 39] Moreover, α-MoO$_3$ layers support a set of electromagnetic modes across the different RBs.[40, 41] The dispersion relation for the PhP in the high-momentum approximation is given by:[39, 40]

$$\sqrt{k_x^2 + k_y^2} - \frac{\rho(k_x,k_y,\omega)}{d}\left[\text{atan}\left(\frac{\rho(k_x,k_y,\omega)\varepsilon_1}{\varepsilon_z}\right) + \text{atan}\left(\frac{\rho(k_x,k_y,\omega)\varepsilon_2}{\varepsilon_z}\right) + \pi l\right] = 0, \quad l = 0,1,2 \ldots \quad (1)$$

where $\rho(k_x, k_y, \omega) = \sqrt{\frac{\varepsilon_z(k_x^2+k_y^2)}{\varepsilon_x k_x^2+\varepsilon_y k_y^2}}$. Here, $k_x$ and $k_y$ are the complex-valued projections of the wavevector onto the $x$- and $y$-axes, respectively; $\varepsilon_1$ and $\varepsilon_2$ are the dielectric permittivities of the superstrate and substrate; and $\hat{\varepsilon} = \text{diag}(\varepsilon_x, \varepsilon_y, \varepsilon_z)$ is the dielectric permittivity of the α-MoO$_3$ layer. All the dielectric permittivities are in general functions of $\omega$. In this Letter, we focus on the fundamental M0 mode ($l = 0$ in Eq. (1)), which exhibits the smallest wave vector and longest propagation length among the available modes. We align the $x$-, $y$-, and $z$-axes with the [100], [001], and [010] crystallographic directions of α-MoO$_3$, respectively, whereas the $z$-axis is orthogonal to the surface of the layer. For our analysis, we consider the RB from 850 to 960 cm$^{-1}$, within which hyperbolic phonon polaritons (HPhPs) propagate within the angular sector centered along the [100] direction (further details in the Supporting Information, Section 1).

In the absence of losses, for a fixed frequency $\omega$, Eq. (1) defines the IFC in a two-dimensional space of the in-plane wavevector components, $\mathbf{k} = (k_x, k_y)^\text{T}$ (for simplicity, we will refer to the real and imaginary in-plane components of the wavevector as $\mathbf{k}'$ and $\mathbf{k}''$, respectively). In the presence of losses, however, $k_x$ and $k_y$ become complex-valued, making Eq. (1), which relates two complex numbers, equivalent to two equations involving four real numbers: $k_x'$, $k_x''$, $k_y'$, and $k_y''$. Consequently, for any fixed direction of $\mathbf{k}'$ (i.e. for any ratio between $k_x'$ and $k_y'$), Eq. (1)

admits an infinite number of solutions — a continuum— corresponding to different magnitudes and directions of $\mathbf{k}''$.[39] Importantly, the dispersion relation does not intrinsically establish any connection between the direction of $\mathbf{k}'$ and $\mathbf{k}''$. Nevertheless, only specific combinations of $\mathbf{k}'$ and $\mathbf{k}''$ can correspond to physically observable polaritonic waves. These combinations depend critically on the characteristics of the excitation source and the experimental setup used for their observation. We consider polaritons in an α-MoO$_3$ crystal layer launched by a point-like source. While the latter does not directly produce plane polariton waves, it generates polaritons propagating in any direction, providing a comprehensive scenario to study their behavior. Specifically, we use a gold rod nanoantenna, ($l$ = 3.4 μm in length, $w$ = 290 nm in width, and $t$ = 40 nm in thickness), on a 225-nm-thick α-MoO$_3$ layer over a SiO$_2$ substrate[42] (Figure 2a; $\varepsilon_1$ = 1 and $\varepsilon_2$ taken from [43]). The antenna is aligned with the $x$-axis, and an external s-polarized illuminating wave, with the electric field parallel to the longest side of the antenna, excites hot spots at its apexes. These hotspots act as effective point sources, launching PhPs within a sector defined by the IFCs asymptotes.[24, 44, 45] As shown in the Supporting Information, this field closely matches that from a point source and is robust to antenna orientation; moreover, the elongated rod-like antenna optimal for our purpose. To visualize the polariton propagation, we carry out scattering-type near-field optical microscopy (s-SNOM; see the Supporting Information, Section 3)[31, 46-48] at $\omega$ = 925.9 cm$^{-1}$. PhPs in α-MoO$_3$ manifest as hyperbolic fringes, as shown in Figure 2b (top panel), with their amplitude decaying away from the source due to both material losses and geometrical spreading. To corroborate the origin of the fringes, we compare the measured near-field map (bottom panel) with the vertical field distribution $E_z(x, y)$ obtained from finite-element-method simulations (top panel). The excellent agreement between the simulated and experimental results verifies that the near-field images are dominated by the $E_z$ of PhPs launched by the antenna.

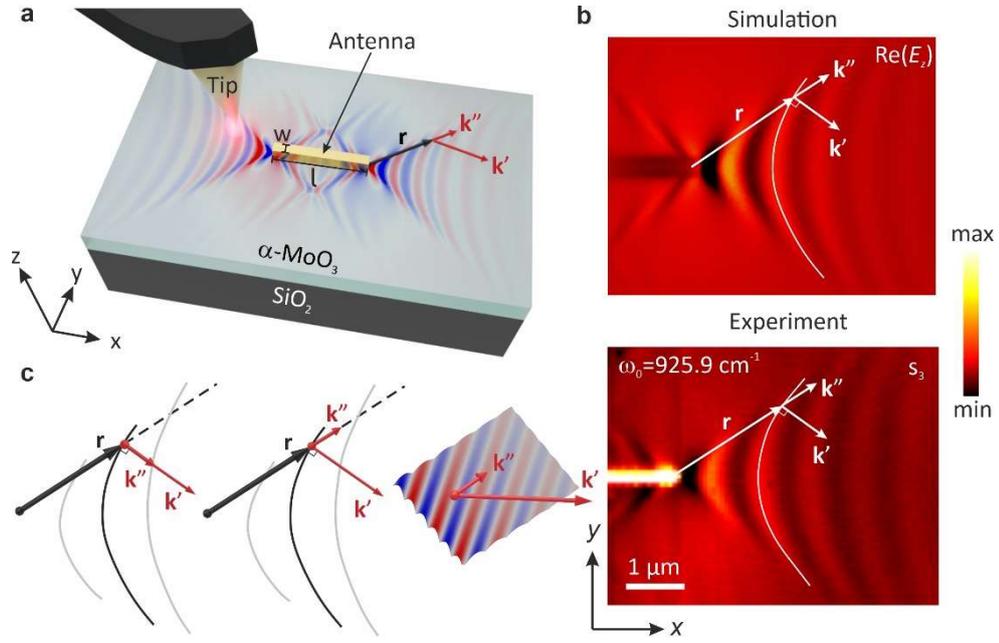

**Figure 2.** Excitation of anisotropic polaritons with non-collinear **k′** and **k″** by an optical antenna and their imaging by s-SNOM. (a) Schematics of a sample in which a gold rod (with length = 3.4 μm, width = 290 nm and thickness = 40 nm) acting as a resonant nanoantenna launches anisotropic polaritons in an α-MoO$_3$ layer (with thickness = 225 nm) placed on top of SiO$_2$ (with dielectric constants ε$_1$=1 and ε$_2$ taken from [43] for simulations). Arrows show the real and imaginary parts of the wavevector, **k′** and **k″**, of the polaritonic mode propagating along the observation vector, **r**. (b) Top panel: numerically simulated real part of the polariton electric field, Re[$E_z(x,y)$], excited in the configuration shown in (a). The field distribution is extracted at a height $h = 100$ nm over the α-MoO$_3$ surface and the nanoantenna is illuminated normally by a plane wave. Bottom panel: s-SNOM near-field amplitude, s$_3$, of anisotropic polaritons excited in the same configuration as in (a) and the top panel. (c) The decay direction **k″** of strongly anisotropic polaritons can be interpreted as either parallel to **k′** (left panel) or to the observation vector **r** (middle panel); the right panel represents an example of the single plane with non-collinear **k′** and **k″**.

To better understand the near-field images and the underlying physics, recall that the polariton spatial field distribution can be represented by a Fourier integral – a continuum sum of plane waves with real **k**. Plane waves with complex **k** cannot form a complete basis for representing the field across all space due to the exponential divergence of their amplitudes in the direction opposite to $\mathbf{k}''$. Nevertheless, plane waves with complex wavevectors satisfying Eq. (1) can describe fields in specific regions.[39] In this approach, $\mathbf{k}'$ is defined by the phase gradient and the amplitude of plane waves includes a geometrical decay factor accounting for the spatial spreading of energy. To determine the corresponding $\mathbf{k}''$ for a given direction of $\mathbf{k}'$, we assume that the wave with the smallest magnitude of $\mathbf{k}''$ (i.e., the smallest losses) dominates the electromagnetic field at points sufficiently far from the source. This minimal value of $\mathbf{k}''$ can be directly obtained from Eq. (1), as explained in detail in the Supporting Information, Section 4. Specifically, for each direction of $\mathbf{k}'$, which characterizes the wavefront at a given observation point, we obtain the direction of the $\mathbf{k}''$ by minimizing its magnitude. Our analysis reveals that the $\mathbf{k}''$ obtained through this minimization procedure is parallel to both the group velocity[39] and to the radius vector, $r$, and therefore, perpendicular to the real IFC (see the Supporting Information, Section 5). This leads to the natural, yet previously overlooked, conclusion that the field along $r$ can be represented by a plane wave exponentially decaying towards the observation direction ($\mathbf{k}''||r$), with $\mathbf{k}'$ not necessarily aligned with $r$. Importantly, the procedure detailed in Section 4 of the Supporting Information is independent of the specific form of the dispersion equation, exemplified here by Eq. (1). Instead, the presented approach is broadly applicable to various types of waves propagating in two dimensions (along a surface or a layer).

Our analysis enables the direct extraction of both $\mathbf{k}'$ and $\mathbf{k}''$ from a near-field image. In both experimental (Fig. 2b, bottom panel) and the numerically simulated (Fig. 2b, top panel) images, as stated above, the electric field distribution along $\mathbf{r}$ can be interpreted as that of a wave with $\mathbf{k}''$ parallel to $\mathbf{r}$, and $\mathbf{k}'$ orthogonal to the wavefront (the curve of a constant phase) at the intersection point with $\mathbf{r}$, as shown in Fig. 2b. Once $\mathbf{k}'$ and $\mathbf{k}''$ directions are known, their magnitudes are obtained by fitting the near-field profile. Specifically, we fit the complex signal $\sigma_n$ along a line at angle $\theta$ to the $x$-axis, using a simple complex-valued function to approximate exponentially decaying waves with a linear background: [35, 49]

$$E_z(r) = c_1 + c_2 r + \frac{A}{\sqrt{r-r_0}} e^{ik'_r r - k''_r r}, \qquad (2)$$

where $c_1$ and $c_2$ are complex-valued fitting parameters that compensate the linear background signal present in the measured data; $k'_r$ and $k''_r$ are the projections of the real and imaginary components of the wavevector along the direction of **r**, which determine the field oscillations and exponential decay, respectively. The constant $A$ represents the complex-valued amplitude of the signal, and $\sqrt{r-r_0}$ is the geometrical decay factor associated to the cylindrical spreading of the wave from the "effective" source position, $r_0$. The cylindrical decay factor is natural for isotropic polaritons from a point source. Although anisotropic polaritons decay direction-dependently and can approach constant factors in the case of canalization, [19, 23] we use an angle-independent geometrical factor, $\sqrt{r-r_0}$, here because it holds along the principal axis, [19] matches analytical point-dipole expressions, [31] and fits our data well in all directions. Figure 3a illustrates the least-square fitting of the measured complex near-field signal profile with $\theta = 15°$ using the function given by Eq. (2). The fitting for all other profiles is provided in the Supporting Information, Section 6. The inset to Figure 3a provides an alternative representation of the complex-valued signal profile as a parametric curve in the complex plane. [9, 49, 50]

The projections $k'_r$ and $k''_r$ are related to the absolute values of **k'** and **k''** as $k'_r = k' \cos(\theta - \theta')$, and $k''_r = k'' \cos(\theta - \theta'')$, where $\theta$, $\theta'$, and $\theta''$ are the angles between the $x$-axis and the **r**, **k'**, and **k''**, respectively. In this representation, the condition **k''** ∥ **r** reads as $\theta'' = \theta$, providig the minimal value of $k''$ (Figure 2c right panel). On the other hand, the condition $\theta'' = \theta'$ describes the "naive" assumption in which **k'**∥ **k''** (Figure 2c left panel). By repeating the field cross-section fitting for the whole set of angles, we extract the angular dependences of $k'_r$ and $k''_r$ and reconstruct the IFCs of the HPhPs from the measurements under both assumptions: $\theta'' = \theta$ and $\theta'' = \theta'$. Interestingly, Figure 3b demonstrates an excellent agreement between the IFCs extracted from the near-field image and those calculated using Eq. (1) for both cases. This agreement illustrates that only the projection of $\mathbf{k}''$ on the propagation direction, $\mathbf{r}$ (and therefore, orthogonal to the IFC) determines the PhP decay. In contrast, the total magnitude of $\mathbf{k}''$ influences the out-of-plane field distribution, as detailed below. To illustrate this statement, we simulate the field generated by a point dipole in z-r plane, with $\mathbf{r}$ forming an angle with the *x*-axis (we take 54° for illustrative purposes) at a fixed frequency $\omega = 950$ cm$^{-1}$, as shown in Figure 4a and 4d. We compare $E_z$ extracted from this simulation with two analytically calculated $E_z$ profiles of waveguiding PhP modes (plane waves) with $\mathbf{k}''$ parallel to $\mathbf{r}$ (Figure 4b), and with $\mathbf{k}''$ parallel to the $\mathbf{k}'$ (Figure 4c). While the distribution of $E_z$ of the PhP mode with $\mathbf{k}'' \parallel \mathbf{r}$ closely resembles the dipole-generated field, the mode with $\mathbf{k}'' \parallel \mathbf{k}'$ rather displays beveled planes of constant phase, resembling ghost polariton waves recently reported [17]. Importantly, for large momenta $k$ we have $k_z \approx ik$, so that $k'$ and $k''$ govern the out-of-plane decay and propagating phase respectively. This implies that the total value of in-plane momentum indeed significantly impacts the out-of-plane field behavior. For a detailed comparison of the out-of-plane field distributions, we plotted Re($E_z$) in Figures 4d and 4e along the horizontal and vertical directions (red and blue dashed lines in Figures 4a-4c, respectively). Consistent with our analysis, Figure 4e demonstrates that the field calculated assuming $\mathbf{k}' \parallel \mathbf{k}''$ exhibits oscillations along the vertical direction, while the field $E_z$ calculated assuming $\mathbf{k}'' \parallel \mathbf{r}$ decays exponentially without any oscillation. The prominent out-of-plane oscillation corresponds to an additional phase gradient along the *z*-direction, resulting in a phase shift between fields calculated under the two above assumptions at some distance from the surface. This phase shift is also evident in Figure 4f, which shows the field distribution 100 nm below the layer. Our analysis highlights the importance of a proper calculation of the imaginary part of the wavevector. While the experimental demonstration of out-of-plane field distribution remains challenging due to limitations in phase measurement techniques across the sample, our results establish a theoretical framework that bridges the gap between simulations and experimentally observable in-plane phenomena. This framework is particularly relevant for systems where precise phase control is essential, such as quantum radiation sources and other nanoscale photonic applications. Moreover, the proper determination of the out-of-plane component of the wavevector can be important when considering the coupling of polaritons in α-MoO$_3$ layer with modes in waveguides or resonators, as well as when numerically calculating the local energy density, $W(x, y, z)$, and the Poynting vector, $S(x, y, z)$, with significant losses in the material. If the misalignment is not taken into account, the error in the mentioned physical quantities can reach the error in determining the out-of-plane component of the wavevector, $|\delta q_z/q_z|$, i.e., in the case considered here, up to 35%.

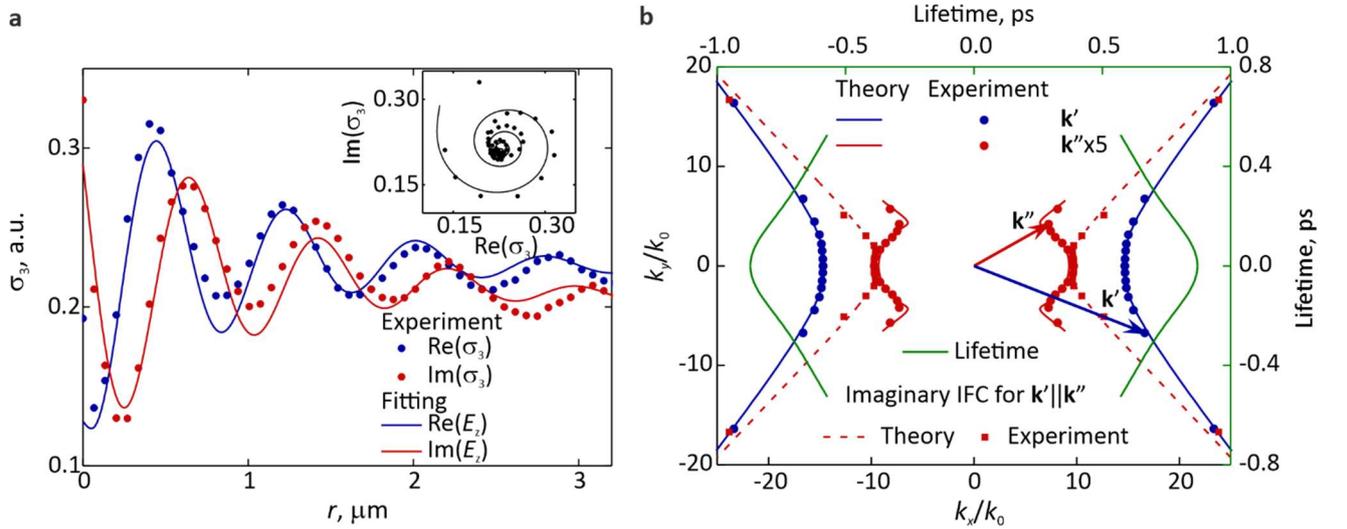

**Figure 3.** Disentangling the complex-valued wavevector **k** of anisotropic polaritons from the near-field measurements. (a) Fitting using Eq. (2) of the real and imaginary parts of the complex-valued near-field signal, $\sigma_3$, along the line oriented at the angle $\theta = 15°$ with respect to the $x$-axis. The inset shows $\sigma_3$ and the fitting function as a parametric data set in the complex plane. (b) Comparison of the real and imaginary IFCs extracted from the fitting of the profiles (dots) and calculated analytically (curves). The dashed curve represents the calculated imaginary IFC assuming **k'** ∥ **k''**. The green curve shows the lifetime in polar coordinates, where the polar angle is for the real part of the wavevector. $k_0$ is the free-space wavevector.

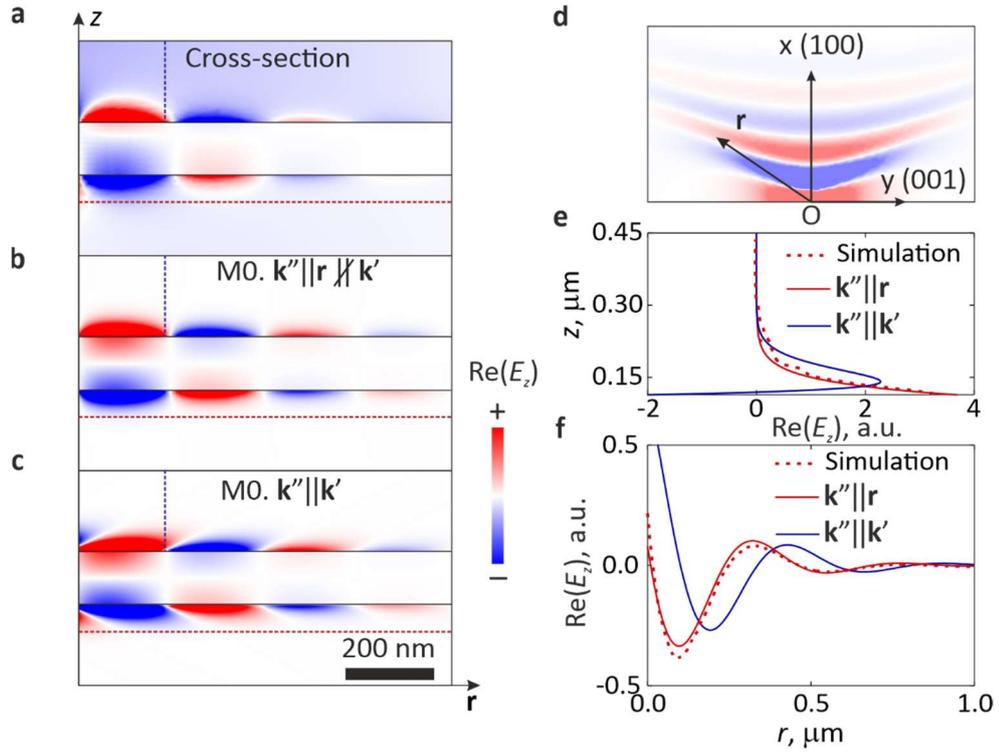

**Figure 4.** Implications of the non-collinearity between **k′** and **k″** for the out-of-plane field distribution. (a) A cross-section of the numerically simulated spatial distribution of Re($E_z$) excited by a vertical point dipole located 500 nm above a 225-nm-thick α-MoO$_3$ layer on a SiO$_2$ substrate at a frequency of 950 cm$^{-1}$. The cross-section is made along the direction forming an angle of 54° with the $x$-axis. (b) analytically calculated field distribution of the M0 polariton mode mostly contributing to the field distribution in panel (a). **k″** is assumed to be parallel to the observation vector, **r**. (c) same as panel (b) for **k′** ∥ **k″**. (d) In-plane distribution of Re($E_z$), where the vector **r** belongs to the cross-section plane ($r, z$) in (a). (e) Re($E_z$) as a function of the $z$-coordinate extracted along the blue dashed lines shown in panels (a)-(c). (f) Re($E_z$) as a function of the $r$-coordinate extracted along the red dashed lines shown in panels (a)-(c).

For a deeper understanding of the properties of the PhP modes in a thin layer, we now discuss key parameters such as the Poynting vector, energy density, and energy velocity. The Poynting vector has been previously used to interpret nanoimaging experiments[7, 8, 31] under the assumption that it is orthogonal to the IFC. However, this orthogonality has been proven only for infinite homogeneous media, [51] and not for layered structures. Using the large-$k$ and thin-layer ($d \ll 1/|k_z|$) approximations, and neglecting losses, we derive the following analytical expression for the time-averaged in-plane Poynting vector of the polariton in an anisotropic layer, $\langle \mathbf{S} \rangle$, which reads (see the Supporting Information. Section 7):

$$\langle \mathbf{S} \rangle = -\frac{c|E_t|^2}{8\pi}\left[\frac{\varepsilon k_0}{k^2}\begin{pmatrix}k_x\\k_y\end{pmatrix} + \frac{2i}{k}\begin{pmatrix}\alpha_x k_x\\\alpha_y k_y\end{pmatrix}\right], \quad (3)$$

where $E_t$ is the amplitude of the in-plane component of the electric field, $\varepsilon = \frac{\varepsilon_1+\varepsilon_2}{2}$, and $\alpha_i = \frac{k_0 d \varepsilon_i}{2i}$ (with $i=x,y,z$) represents the normalized effective conductivity of the layer. [9] We also calculate the energy density distribution of an electromagnetic wave propagating in a thin anisotropic layer:

$$\langle W \rangle = \frac{k_0 dc|E_t|^2}{16\pi k^2}\left(\frac{d\varepsilon_x}{d\omega}k_x^2 + \frac{d\varepsilon_y}{d\omega}k_y^2\right). \quad (4)$$

To obtain the in-plane Poynting vector and energy density given by Eqs. (3) and (4), we integrated the 3D distribution of the Poynting vector and energy density respectively over the $z$-axis. Expressions (3) and (4) allow us to directly find the energy velocity and show that the group and energy velocities in this case are equal to each other:

$$\mathbf{v}_E = \frac{\langle \mathbf{S} \rangle}{\langle W \rangle} = \frac{-2}{\left(\frac{d\varepsilon_x}{d\omega}k_x^2+\frac{d\varepsilon_y}{d\omega}k_y^2\right)}\left[\frac{\varepsilon}{kd}\begin{pmatrix}k_x\\k_y\end{pmatrix} + \begin{pmatrix}\varepsilon_x k_x\\\varepsilon_y k_y\end{pmatrix}\right] = \frac{\partial\omega}{\partial \mathbf{k}} = \mathbf{v}_{gr} \quad (5)$$

The expressions for the Poynting vector and the energy density are crucial tools for describing the properties of electromagnetic modes, and, in particular, calculating parameter such as the lifetime. In the low-losses regime ($\tau \ll \omega$, typically valid for propagating polaritons), the lifetime, $\tau$, explicitly reads as (see the Supporting Information. Section 7):

$$\tau^{-1} = 2\mathbf{k}'' \cdot \mathbf{v}_{gr} \quad (6)$$

Noticeably, only the projection of $\mathbf{k}''$ onto $\mathbf{v}_{gr}$ determines the lifetime. On the other hand, in the regime of small losses, all solutions of Eq. (1) for a certain direction of $\mathbf{k}'$ have the same projection of $\mathbf{k}''$ onto $\mathbf{v}_{gr}$, that is, $\mathbf{k}'' \cdot \mathbf{v}_{gr} = const$ for the entire continuum of solutions of Eq. (1) with a fixed direction of $\mathbf{k}'$ (see the Supporting Information, Section 5). Thus, despite the uncertainty in $\mathbf{k}''$ for a fixed direction of $\mathbf{k}'$, the lifetime is unambiguously determined; in other words, if the lifetime is calculated in the low-loss regime using Eq. (6), it will not depend significantly on the assumption used about the $\mathbf{k}''$ direction. We can now calculate the angular dependence of the lifetime using expressions (5) and (6). Figure 3b shows the lifetime plotted in polar coordinates, where the angle corresponds to the direction of $\mathbf{k}'$ for the mode under consideration. Remarkably, due to angular dependence of $\mathbf{k}'$ and $\mathbf{k}''$, the lifetime varies in the range of 15% over the entire angular range, under the conditions considered.

We have demonstrated, for the first time, the extraction of the imaginary part of the polariton in-plane wavevector in a biaxial anisotropic layer as a function of propagation direction, based on near-field measurements. The experimentally reconstructed IFCs for both real and imaginary components show excellent agreement with theoretically calculated IFCs under the assumption of non-collinearity between $\mathbf{k}'$ and $\mathbf{k}''$. We derived Poynting vector and energy velocity expressions, confirming the equality of group and energy velocities for large-$k$ modes in low-loss conditions. This framework advances interpretation of near-field data and understanding of wave behavior in anisotropic materials, relevant to strong coupling, hyperbolic reflection, refraction, and related effects. To illustrate the practical implications of our theory, we also calculated the angular dependence of the lifetime of in-plane hyperbolic polaritons. Our findings are broadly applicable and can be extended to other anisotropic materials and heterostructures used in advanced nanophotonic systems.

ASSOCIATED CONTENT

**Supporting Information**

The Supporting Information contains additional dispersion graphs and isofrequency curves of polaritons in MoO3 layers; discussion of antenna shape selection; technical details of sample fabrication, near-field measurements, and numerical simulations; a detailed algorithm for calculating the imaginary part of the wave vector using the minimization method; proof of the collinearity of the imaginary part of the wavevector and the group velocity; the fitting all field profiles for which **k′** and **k″** shown in Fig. 3b are extracted; derivation of the expression for the in-plane Poynting vector, energy density, energy velocity, and lifetime.

AUTHOR INFORMATION

**Corresponding Author**

[†]Contact author: pabloalonso@uniovi.es

[‡]Contact author: alexey@dipc.org

**Funding**

The authors acknowledge the Spanish Ministry of Science and Innovation (State Plan for Scientific and Technical Research and Innovation grant numbers PID2022-141304NB-I00 and PID2023-147676NB-I00). K.V.V. received the support of a fellowship from "la Caixa" Foundation (ID 100010434) with the fellowship code LCF/BQ/DI21/11860026. A.Y.N acknowledges the Basque Department of Education (grant PIBA-2023-1-0007). P.A.-G. acknowledges support from the European Research Council under Consolidator grant no. 101044461, TWISTOPTICS.

**Notes**

The authors declare no competing financial interest.

ACKNOWLEDGMENT